\documentclass[letterpaper,twocolumn,english,prl,showpacs]{revtex4}
\usepackage{mathptmx}
\pdfoutput=1
\usepackage[T1]{fontenc}
\usepackage[latin9]{inputenc}
\usepackage{array}
\usepackage{longtable}
\usepackage{amsmath}
\usepackage{graphicx}
\usepackage{amssymb}

\makeatletter

\providecommand{\tabularnewline}{\\}

\@ifundefined{textcolor}{}
{%
 \definecolor{BLACK}{gray}{0}
 \definecolor{WHITE}{gray}{1}
 \definecolor{RED}{rgb}{1,0,0}
 \definecolor{GREEN}{rgb}{0,1,0}
 \definecolor{BLUE}{rgb}{0,0,1}
 \definecolor{CYAN}{cmyk}{1,0,0,0}
 \definecolor{MAGENTA}{cmyk}{0,1,0,0}
 \definecolor{YELLOW}{cmyk}{0,0,1,0}
 }

\makeatother

\usepackage{babel}

\makeatother

\usepackage{sidecap}

\newcommand{\TabBesBeg}{%
 \let\MyTable\table
 \let\MyEndtable\endtable
 \renewenvironment{table}{\begin{SCtable}}{\end{SCtable}}}

\newcommand{\TabBesEnd}{%
 \let\table\MyTable
 \let\endtable\MyEndtable}

\newcommand{\FigBesBeg}{%
 \let\MyFigure\figure
 \let\MyEndfigure\endfigure
 \renewenvironment{figure}{\begin{SCfigure}}{\end{SCfigure}}}

\newcommand{\FigBesEnd}{%
 \let\figure\MyFigure
 \let\endfigure\MyEndfigure}

\sidecaptionvpos{figure}{c}

\makeatother

\usepackage{babel}

\begin{document}

\title{The Fractional Quantum Hall state at $\nu=5/2$ and the Moore-Read
Pfaffian}

\author{M. Storni$^{1}$, R. H. Morf$^{1}$ and S. Das Sarma$^{2}$}

\affiliation{$^{1}$Condensed Matter Theory, Paul Scherrer Institute, CH-5232
Villigen, Switzerland}

\affiliation{$^{2}$Condensed Matter Theory Center, Department of Physics, University
of Maryland, College Park, MD 20742}

\date{{20 December 2008}}
\begin{abstract}
Using exact diagonalization we show that the spin-polarized Coulomb
ground state at $\nu=\frac{5}{2}$ is adiabatically connected with
the Moore-Read wave function for systems with up to 18 electrons on
the surface of a sphere. The ground state is protected by a large
gap for all system sizes studied. Furthermore, varying the Haldane
pseudopotentials $v_{1}$ and $v_{3}$, keeping all others at their
value for Coulomb interaction, energy gap and overlap between ground-
and Moore-Read state form hills whose positions and extent in the
$(v_{1},v_{3})$-plane coincide. We conclude that the physics of the
Coulomb ground state at $\nu=\frac{5}{2}$ is captured by the Moore-Read
state. Such an adiabatic connection is not found at $\nu=\frac{1}{2}$,
unless the width of the interface wave function or Landau level mixing
effects are large enough. Yet, a Moore-Read-phase at $\nu=\frac{1}{2}$
is unlikely in the thermodynamic limit. 
\end{abstract}


\maketitle
One of the most intriguing strongly correlated electronic states discovered
in nature is the even-denominator fractional quantum Hall effect (FQHE)
at the Landau level filling factor $\nu=5/2=2+1/2$ \cite{Willett87},
i.e. at the half-filled second orbital Landau level (LL) of a 2D electron
system. The $5/2$ FQHE cannot be understood within the canonical
hierarchical (Laughlin) theory, since the odd-denominator rule is
a necessity to preserve the Pauli principle. A particularly interesting
proposal by Moore and Read (MR) \cite{MooreRead91} extending Laughlin's
ideas to quantum Hall states at half filling is the {}``Pfaffian''
wave function (wf), characterized by quasiparticle excitations obeying
non-Abelian braiding statistics \cite{NayakRMP08}.

The first numerical study of this wf was carried out by Greiter \textit{et
al.} \cite{Greiter92} who considered it as a candidate for the observed
FQHE at both $\nu=1/2$ and $5/2$. Their calculations done for systems
on the sphere with $N_{el}\le$10 electrons did not allow a determination
of the excitation gap and the difference between $\nu=1/2$ and $5/2$
was not explored in any detail. A first hint at possible adiabatic
continuity (AC) between the MR state and the ground state (GS) of
a two-body model interaction was mentioned briefly in a subsequent
paper by Wen \cite{Wen93}, but limited to a single system size $N_{el}=10$.

%However, the possible AC with the
%Coulomb GS at $\nu=1/2$ and $5/2$ was not explored and the calculation
%was limited to a single system size $N_{el}=10$.

Shortly after its discovery, the $\nu=5/2$ state was studied in a
tilted magnetic field \cite{eisenstein}. Examining the temperature
dependence of the longitudinal resistivity $\rho_{xx}\approx\exp-(\Delta/2k_{B}T)$,
the activation gap $\Delta$ was found to decrease with increasing
tilt angle and the Hall plateau disappeared beyond some critical tilt
angle. These results suggested that the quantized state is at most
partially spin-polarized until at some critical tilt angle the increasing
Zeeman energy produces a phase transition to a gapless polarized state
\cite{questionable}.

This scenario was challenged by one of us \cite{Morf98}: exact diagonalization
results for small systems on a sphere for spin-unpolarized and fully
polarized states at $\nu=5/2$ have shown that the GS is spin polarized
even for vanishing Zeeman energy. Furthermore, the GS for $N_{el}=8$
electrons was found to have substantial overlap with the MR state
although that state is the exact ground state of an unphysical short-range
three-body interaction Hamiltonian. Subsequent theoretical \cite{Rezayi00,Morf03}
and experimental \cite{pan} studies yielded results consistent with
these ideas.

These findings led Das Sarma et al. \cite{DasSarma05} to propose
the use of the $\nu=5/2$ FQH state for the realization of non-Abelian
topological qubits which, they argued, would permit fault tolerant
and robust quantum computation. Their proposal prompted a great surge
of activity \cite{NayakRMP08} to further elucidate the nature of
the $5/2$ FQHE both theoretically \cite{Peterson08,Wan,Feiguin,moller}
and experimentally \cite{Radu08,Dolev08,Dean08}. However, whether
the FQHE at $\nu=5/2$ observed in experiments has the properties
of the non-Abelian MR state remains an open problem, especially since
the relevance of the {}``Pfaffian'' state at $\nu=5/2$ has been
questioned by \cite{Toke}: In their exact diagonalization studies
of quasiholes (qhs), they only observed qhs with charge e/2, while
the qhs in the MR state are predicted to have charge e/4 \cite{qh}.

\begin{figure*}[t]
\includegraphics[width=15.5cm]{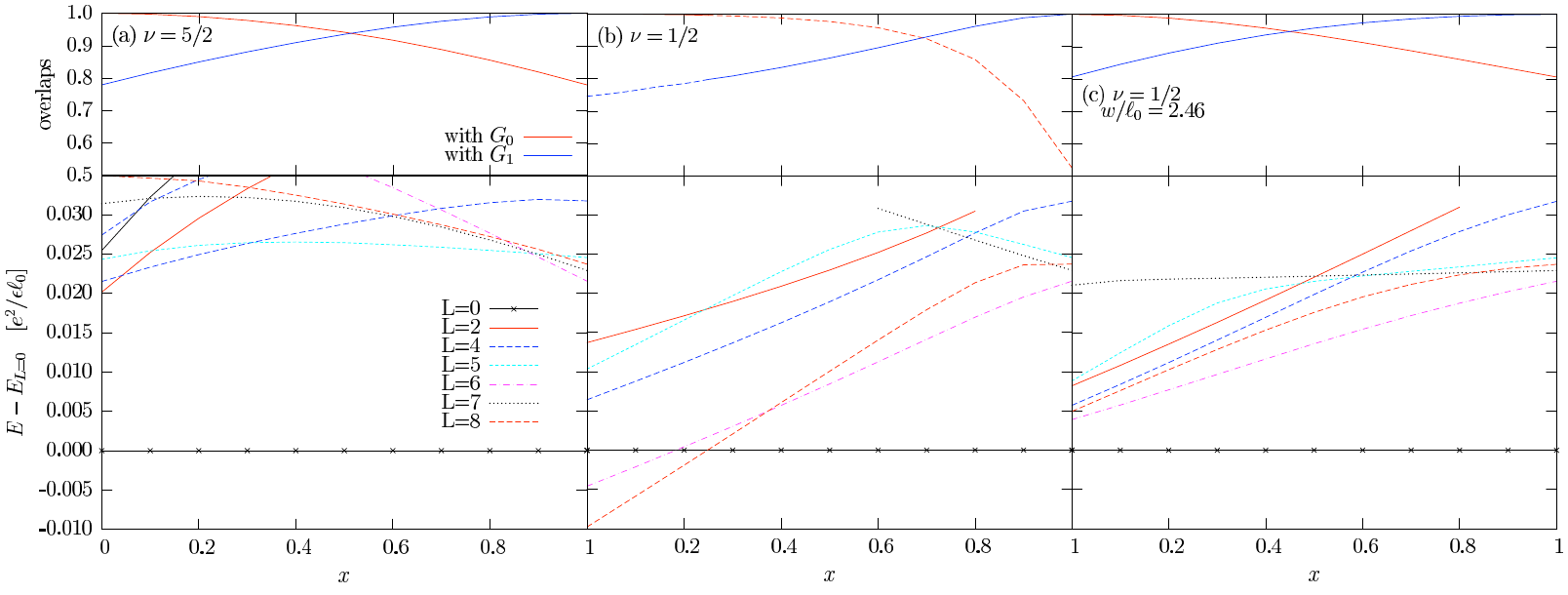}

\caption{$N_{el}=16$, $N_{\phi}=29$: low-lying energy spectra (the lowest
$L=0$-state is the reference) and overlaps $\langle G_{1}|G_{x}\rangle$
and $\langle G_{0}|G_{x}\rangle$ as function of the interaction parameter
$x$. Figure (a) $\nu=5/2$, (b) $\nu=1/2$, (c) $\nu=1/2$ for finite
width $w/\ell_{0}=2.46$. In (b) overlap curves are shown as dashed
lines for $x$-values for which the state $|G_{x}\rangle$ is not
the GS (color online).}

\end{figure*}

In this Letter we provide theoretical evidence, using state of the
art exact diagonalization, that the MR wf and the spin-polarized $\nu=5/2$
FQH state belong to the same universality class \cite{univ}. Following
the pioneering work by Haldane and Rezayi \cite{Haldane} who established
that the GS at $\nu=1/3$ is in the universality class of the 1/3-Laughlin
state, we adiabatically change the electron interaction by interpolating
between the three-body interaction $V_{3b}$, for which the Pfaffian
wf is the unique GS, and the Coulomb interaction $V_{C}$ and follow
the evolution of GS and energy spectrum by exact numerical diagonalization.

For all even system sizes examined ($N_{el}\le18$) we observe AC
of the GS and no indication of a decrease of the gap for interactions
interpolating between $V_{C}$ and $V_{3b}$, thus implying AC between
the spin-polarized $5/2$ state and the MR state in the thermodynamic
limit.

A related study is mentioned in a recent paper by Möller and Simon
\cite{moller}. They report that in systems of $12$, $14$ and $16$
electrons they see no gap closing when interpolating the interaction
between $V_{3b}$ and a particular type of two-body interaction near
the Coulomb interaction, but supposedly {}``in the weak pairing phase''
\cite{moller}. Contrary to our present Letter, no details are given
and the difference between $\nu=1/2$ and $5/2$ is not discussed.

%In our present work, we show and discuss the results for gaps and
%overlaps and present direct evidence of AC for $N=16$ electrons on
%a sphere.

In addition, we systematically vary the two-body interaction, by using
the Haldane pseudopotentials determining the pairwise interaction
among the electrons, and construct a phase diagram which elucidates
the difference between $\nu=1/2$ and $\nu=5/2$ and allows a discussion
of the influence of experimental parameters and Landau level mixing
on the nature of the state. In this phase diagram the region that
corresponds to the gapped phase coincides with the region where the
exact numerical GS has a large overlap with the MR wf, further corroborating
the AC between the realistic GS and the MR state.

%Furthermore, the quantum phase diagram of the incompressible 5/2 FQH
%tate, as obtained from the finite excitation gap calculated by using
%the Haldane pseudopotentials determining the pairwise interaction
%among the electrons, coincides with our direct finding of a large
%overlap between the exact numerical ground state and the MR wf, further
%corroborating the AC between the realistic ground
%state and non-Abelian Moore-Read state \cite{MooreRead91}.

The possibility of MR FQHE at $\nu=1/2$ has first been discussed
in \cite{Greiter92}, but the consensus \cite{He93,Suen94} seems
to be that the $\nu=1/2$ state in a single-layer 2D system is unlikely
to be the MR state. In this Letter, we find that under specific conditions
(e.g. the thickness of the layer should be a few magnetic lengths
wide) the $\nu=1/2$ state can become adiabatically connected to the
MR wf for small systems. Yet for increasing system size, the gap decreases
in a way that it is doubtful that the Moore-Read state can occur in
single-layer systems at $\nu=1/2$.

%We have discovered in our numerical results the possible existence of a hitherto unknown 
%incompressible FQHE at nu=5/2 at low values of the pseudopotential parameter V1, which is not 
%adiabatically connected to the MR pfaffian state.  We speculate that this new 5/2 FQH state may be 
%he elusive strong-pairing Abelian p-wave state mentioned as a possible candidate state earlier in 
%the literature [Read-Green]. To the best of our knowledge this is the first theoretical finding of 
%n incompressible $5/2$ FQHE, which is not adiabatically connected to the MR pfaffian state.

In the following we study the low-lying energy spectra of fully spin-polarized
systems with up to $18$ electrons in the half filled first ($\nu=1/2$)
and second ($\nu=5/2$) LL. If the ground state (GS) has angular momentum
$L=0$ (and is therefore rotation-invariant) we consider as {}``FQH
gap'' the energy difference between the GS and the first excited
state, although the real gap corresponding to an exciton where quasiparticle
and quasihole are infinitely separated will be somewhat larger \cite{morf_dam_das}.

We use Haldane's spherical geometry \cite{Haldane}, in which for
a half filled LL the particle number $N_{el}$ and the number of flux
quanta $N_{\phi}$ are related by $N_{\phi}=2N_{el}-S$. Here, the
{}``shift'' $S$ is a topological quantum number \cite{WenZee92}
and depends on the particular FQH state: for the MR state $S=3$.
We consider particle interactions of the form,\begin{equation}
V=(1-x)V_{2b}+xV_{3b},\label{eq:1}\end{equation}
 with $0\le x\le1$, interpolating between a generic 2-body potential
$V_{2b}$ and the 3-body interaction $V_{3b}$ for which the MR wf
is an exact GS \cite{Wen93}: \begin{equation}
V_{3b}=\frac{A}{N_{el}^{5}}\sum_{i<j<k}^{N_{el}}S_{ijk}\left\{ \Delta_{j}\delta(i-j)\Delta_{k}^{2}\delta(i-k)\right\} ,\label{eq:2}\end{equation}
 where $\delta(i-j)$ is the $\delta$-function in the separation
of particles $i$ and $j$, $S_{ijk}$ denotes symmetrization over
the permutations within the triplet $(ijk)$. We choose the constant
$A$ such that $V_{3b}$ generates the approximately same gap as the
Coulomb interaction in the second LL. Note that $V_{3b}$ leads to
a GS energy which is extensive for any $\nu>1/2$. By projection of
$V_{3b}$ on a LL the singularities of the $\delta$ functions are
regularized. The 2-body interaction $V_{2b}$ can be written as \begin{equation}
V_{2b}=\sum_{m=0}^{N_{\Phi}}v_{m}\sum_{1\le i<j\le N_{el}}P_{m}(ij),\label{eq:3}\end{equation}
 where $P_{m}(ij)$ is the projector on the states in which particles
$i$ and $j$ have relative angular momentum $m\hbar$ and the pseudopotential
$v_{m}$ \cite{Haldane} is the corresponding LL-dependent interaction
energy.

For $V_{2b}$ we first take the Coulomb interaction $V_{C}$ of point
particles and study the evolution of the ground state $|G_{x}\rangle$
and lowest energy excited states as function of the parameter $x$.
In Fig.~1 we show the energies (lower panel) and overlaps $\langle G_{0}|G_{x}\rangle$
and $\langle G_{1}|G_{x}\rangle$ (upper panel), where $|G_{1}\rangle$
is the MR wf (unique GS for $V_{3b}$) and $|G_{0}\rangle$ is the
Coulomb GS. Energies are measured in the usual units $e^{2}/\epsilon\ell_{0}$,
where $\ell_{0}=\sqrt{\hbar c/eB}$ is the magnetic length.

\begin{table*}
\begin{longtable}{>{\raggedright}b{12cm}>{\raggedright}p{5.5cm}}
\includegraphics[width=11.6cm]{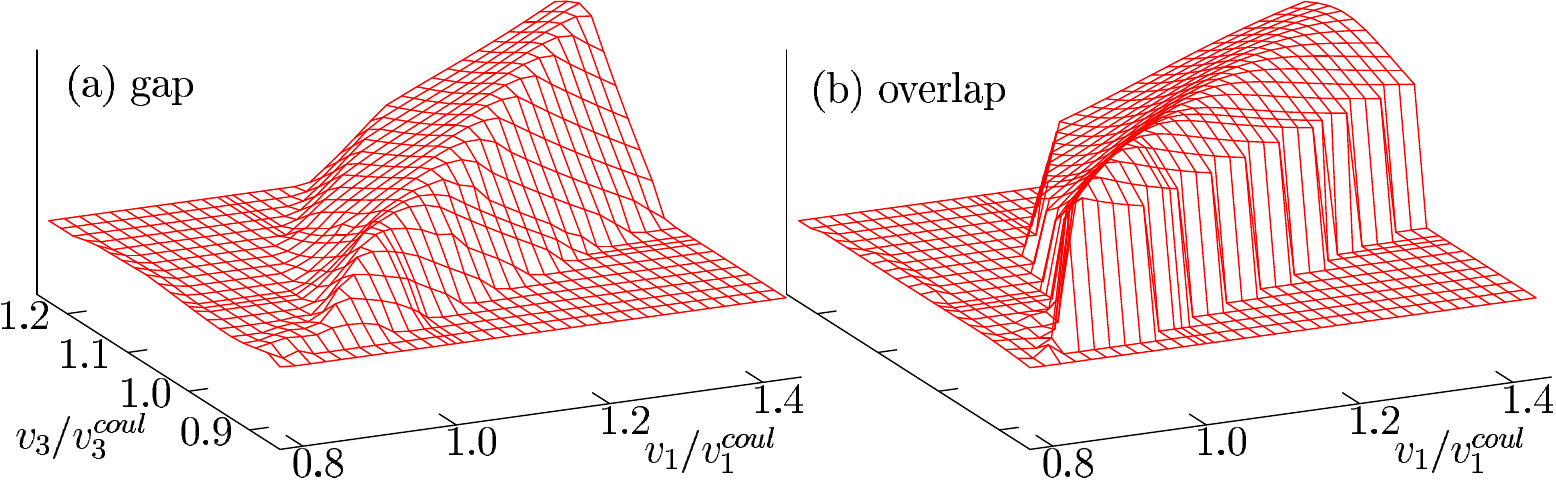}

\includegraphics[width=5.8cm]{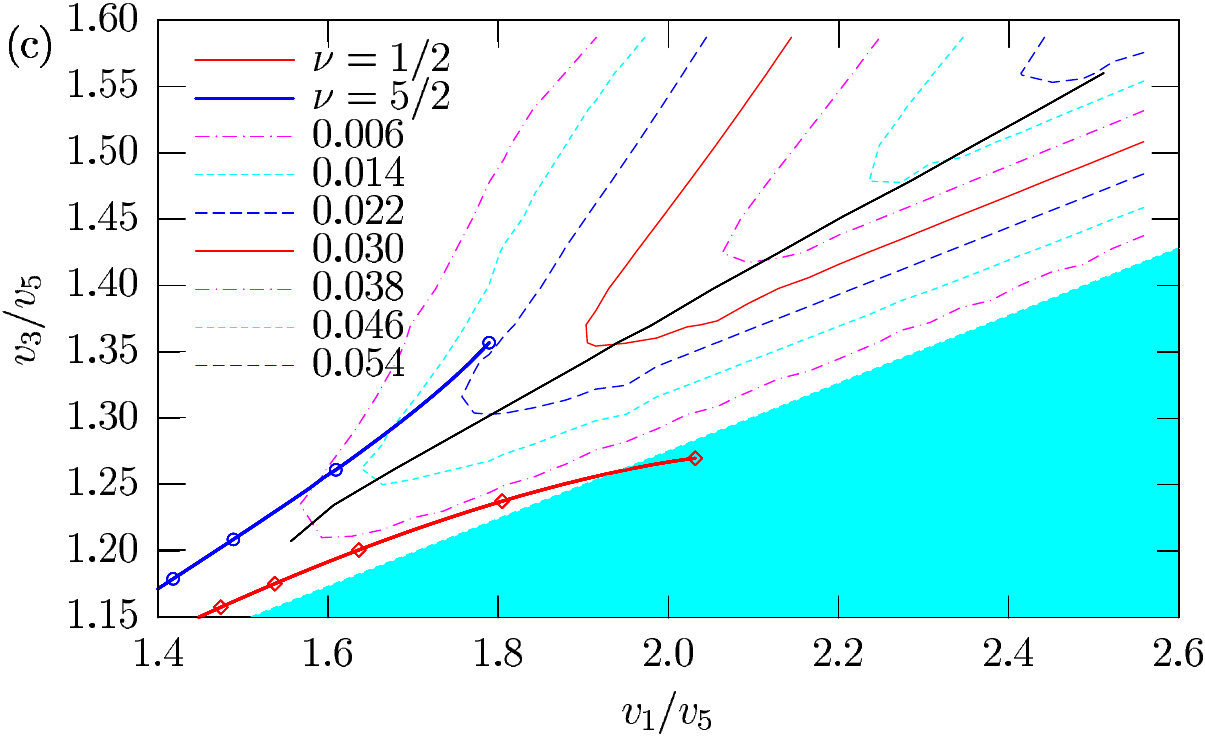}\includegraphics[width=5.8cm]{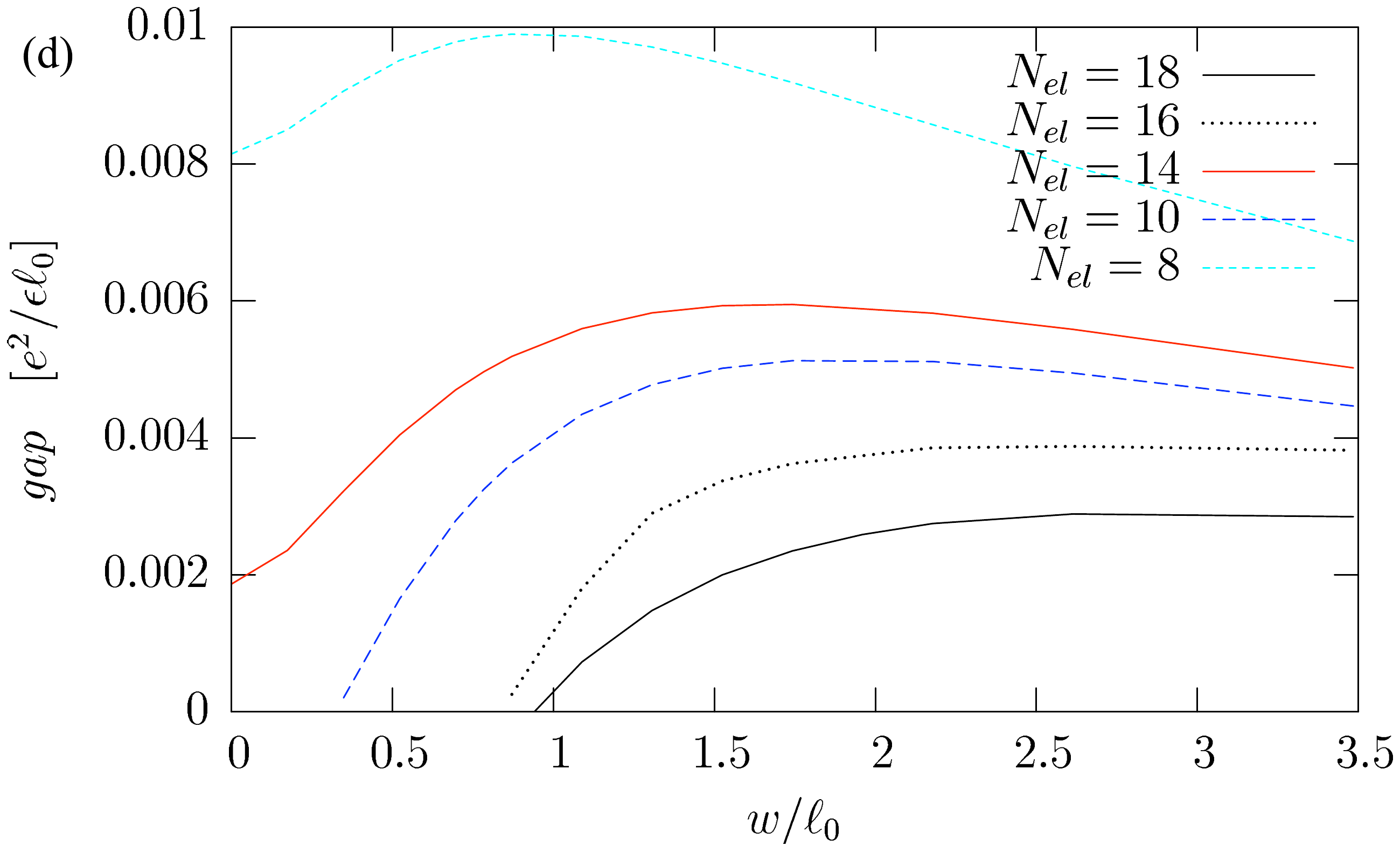} & \includegraphics[width=5.4cm]{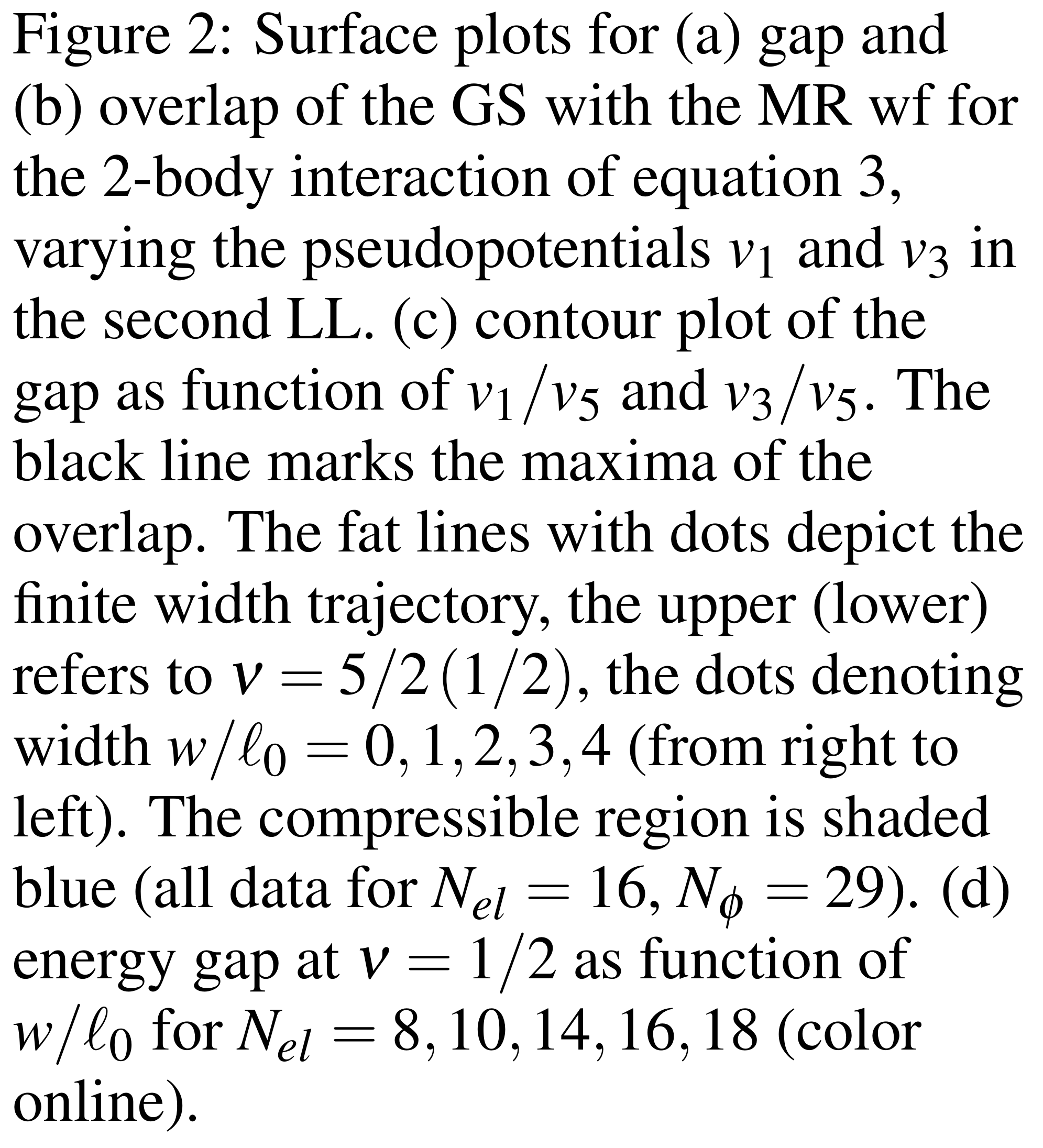}\tabularnewline
\end{longtable}
\end{table*}

Fig.~1(a) shows the results in the second LL: Varying $x$ from $0$
to $1$ the GS has always angular momentum $L=0$ and the lowest excitation
energy has a weak maximum near $x=1/2$ with values 0.0257, 0.0300,
0.0248, 0.0263, 0.0264, 0.0240 for $N_{el}=8,10,\,12,\,14,\,16$ and
$18$, respectively. Furthermore, as $x$ is lowered, the overlap
with the MR wf slowly decreases reaching a fairly high value at the
Coulomb point ($\approx0.78$ for $N_{el}=16$ electrons). At the
Coulomb point ($x=0)$ and at the Moore-Read point ($x=1$), we have
evidence of a finite gap in the thermodynamic limit (i.e. with infinite
quasiparticle-quasihole separation). The gap in the thermodynamic
limit has been calculated in \cite{Morf98,morf_dam_das} for Coulomb
interaction, while its value for the MR state, with $A=0.0005$, is
about $0.024$. Since our calculated gaps for $0\le x\le1$ and for
all system sizes are never smaller than at the two end points $x=1$
and $x=0$, we may expect AC at $\nu=5/2$ between Coulomb GS and
MR-state in the thermodynamic limit.

The structure of the excitation spectrum (particularly its $L$-dependence)
depends on $x$. This must be expected: the interaction between quasiparticles
and quasiholes will cause some reordering of excited states. A similar
observation was made studying the $\nu=1/3$ state \cite{Haldane}.

In the lowest LL at $\nu=1/2$, (see Fig.~1(b)) the situation is
different: As $x$ is reduced, at $x=x_{c}$ we observe a phase transition
to compressible GS and for $0\le x\le x_{c}$ the GS has angular momentum
$L>0,$ and thus vanishing overlap with the Paffian and no AC to the
MR state. The situation changes when the finite width of the wave
function in the direction perpendicular to the 2D electron system
is taken into account. Our results of Fig.~1(c) reveal that a small
gap opens down to $x=0$ and the overlaps are comparable in size or
even larger than in the second LL. Thus the finite width induces adiabiatic
connection between MR wf and Coulomb GS in the lowest LL. %Interestingly, we note that in this case the
%excitation spectrum is {}``better'' behaved than in the second LL:
%the order of low-lying energy levels remains almost the same for $0\le x\le1$.
In analogy, we also look at the effect of a finite width in the second
LL: In agreement with \cite{Peterson08}, we obtain a decrease of
the Coulomb gap, together with an increase of the overlap between
MR wf and Coulomb GS. We note LL mixing effects can be accounted for
by an effective width in the range $1\lesssim w/\ell_{0}\lesssim6$,
depending on the cyclotron energy (and electron density) \cite{Morf03,nuebler2009}.

%\begin{figure}[b]
% \includegraphics[width=8cm]{figure_2}
%\caption{Coulomb gaps in the lowest LL as function of the interfacial wf width
%$w/\ell_{0}$ for different system sizes. The inset shows the maximum
%of each curve as function of $1/N_{el}$.}
%\end{figure}

%In the inset

To study in detail the finite width effect and the difference between
first and second LL we vary the 2-body interaction $V_{2b}$ by changing
the pseudopotentials $v_{1}$ and $v_{3}$ (Equation \ref{eq:3})
and keeping all other $v_{m}$ at their Coulomb values (in a given
LL). The values of $v_{i}$ encode the dependence of the interaction
on sample characteristics, like the width of the 2D layer and the
electron density.

In Fig.~2(a) we plot the gap as function of $v_{1}/v_{1}^{Coul}$
and $v_{3}/v_{3}^{Coul}$ for $16$ particles in the second LL, where
$v_{i}^{Coul}$ are the Coulomb values of the pseudopotentials. In
Fig.~2(b) we do the same for the overlap of the GS with the MR wf.
In both cases we find hills with ridges whose positions are close
to a straight line given by an approximately fixed $v_{3}/v_{1}$-ratio:
the gap ridge increases approximately linearly along this line, the
overlap ridge rises quickly, reaching values well above 0.9 even for
the largest systems ($N_{el}\le16$). Remarkably, these two hills
are congruent in position for a given system size, while their extent
and shape show onlylittle system size dependence. The two hills of
gap and overlap thus belong together and are a manifestation of the
{}``MR phase''; below the hills, for smaller $v_{3}$, we find a
compressible phase.

We also note that, if we plot the gaps and overlaps as functions of
$y_{1}=v_{1}/v_{5}$ and $y_{3}=v_{3}/v_{5}$, the resulting plots
for $\nu=1/2$ and $\nu=5/2$ are quite similar, the differences being
of the same magnitude as those due to finite size effects. This results
from the fact that the higher order $v_{m}$'s change only little
when going from the lowest to the second LL. In Fig.~2(c) we summarize
our results for both LL in the $(y_{1},y_{3})$-plane: the gap contour
plot shows the incompressible region, in addition the black line marks
the top of the overlap ridge; the shaded (blue) area is the compressible
region.

Now looking at the finite width $y_{3}$($y_{1})$-trajectories we
can view the above results in a new light: for $\nu=5/2$ (thick line
marked with 4 dots) the Coulomb point is on the {}``safe side''
of the MR gap ridge, with a consistent gap and a high overlap with
the MR wf; increasing the thickness of the system the overlap grows
somewhat, as the finite width trajectory approaches the crest of the
ridge, while the gap decreases. For $\nu=1/2$ (thick line marked
with 5 dots) the situation is very different: the Coulomb point is
on the other side of the MR ridge, near the line of the phase transition,
for some system sizes in the gapped region, for others already in
the compressible phase. We thus conclude that the MR-phase is so close
to the compressible domain that a definite prediction of its existence
in the thermodynamic limit is not possible and only experiment can
answer.

Indeed, the gaps calculated for $\nu=1/2$ for finite width are small
(Figure 2d) and show a marked, although non-monotonic, decrease with
increasing system size ($N_{el}\leq18$) while the layer width at
which the gap opens increases with system size. It is unlikely that
a gap survives in the thermodynamic limit for any layer or quantum
well width supporting a single layer system \cite{HLR}. This proves
the importance of careful studies of the system size dependence for
valid conclusions about the existence of FQH states.

As a test of our methods in discriminating the MR phase from Abelian
FQH phases we studied the system with $N_{el}=12$ and $N_{\phi}=2N_{el}-3=21$,
which is {}``aliased'' with the hierarchical $2/5$-state of 10
holes, $N_{\phi}=\frac{5}{2}N_{holes}-4=21$ \cite{Ambrumenil89}.
Indeed, our results for the second LL show AC between the Coulomb
GS and the MR state. However, in the lowest LL as the interaction
is varied from pure three-body to Coulomb, the gap increases linearly
while the overlap of the MR wf with the GS decreases strongly and
its largest overlap is with a high-lying $L=0$ state ($\Delta E\approx0.128$):
we have entered the Abelian hierarchical phase. The phase transition,
as the interaction is varied, from the non-Abelian MR phase to the
Abelian hierarchy phase is signalled by a significant and sharp decrease
of the overlap between the GS and the MR state. To identify the universality
class of a FQH state, AC is thus only a necessary condition, one must
also study the overlap between GS and prototype FQH state as well
as its system size dependence.

Finally, we address the choice of the shift $S=3$: Three important
features characterize states at $S=3$: (i) the GS at 5/2 has angular
momentum $L=0$ for all even system sizes $N_{el}\le20$ explored
by us. (ii) the excitation gap shows a smooth size dependence as expected
for a FQH state \cite{Morf98,morf_dam_das}. (iii) low energy states
at $S'\ne3$ can be consistently identified as states with $N_{qp}=\pm2(S'-3)$
quasiparticles of charge $\pm e/4$ nucleated in the underlying FQH
state with $S=3$, while the GS has small angular momentum $L=O(N_{el}^{0})$
indicating that quasiparticles with charge $\pm e/4$ are well separated.

We have shown that the polarized GS for Coulomb interaction at $\nu=5/2$
is adiabatically connected to the Moore-Read state for all sizes studied.
If the gap does not close in the thermodynamic limit - we have not
seen any sign that it will - the polarized GS at $\nu=5/2$ has the
characteristics of the MR state. The same may happen in the lowest
LL at $\nu=1/2$: While finite width and LL mixing effects may help
establish a Moore-Read phase, its realization in the thermodynamic
limit remains doubtful.

We acknowledge fruitful discussions with N. d'Ambrumenil, J. Fröhlich
and B.I. Halperin and support by the Swiss National Science Foundation
and the Institute for Theoretical Physics at ETH, Zurich.


\begin{thebibliography}{30}
\bibitem{questionable}For this reason, it was questionable whether
the results of \cite{Greiter92} had any relevance for the observed
$\nu=5/2$ state.
\bibitem{univ} With universality class we mean the class of all states
that are adiabatically connected to a prototypal state.
\end{thebibliography}
\end{document}